\documentclass[preprint,preprintnumbers,nofootinbib]{revtex4}

\usepackage[dvips]{color}
\usepackage{graphicx}
\usepackage{bbm}

\newcommand{\Tr}{{\rm Tr}}
\newcommand{\diag}{{\rm Diag}}
\newcommand{\ev}{{\rm eV}}
\newcommand{\mev}{{\rm MeV}}
\newcommand{\gev}{{\rm GeV}}

\newcommand{\bmx}{\left(\begin{array}}
\newcommand{\emx}{\end{array}\right)}

\begin{document}

\title{Large $\theta_{13}^{}$ from finite quantum corrections in quasi-degenerate neutrino mass spectrum}
\author{
Takeshi Araki$^{a)}$\footnote{araki@ihep.ac.cn} and Chao-Qiang Geng$^{b),c)}$\footnote{geng@phys.nthu.edu.tw}
} 
\affiliation{
$^{a)}$Institute of High Energy Physics, Chinese Academy of Sciences, Beijing 100049, China \\
$^{b)}$Department of Physics, National Tsing Hua University, Hsinchu 300, Taiwan\\
$^{c)}$Physics Division, 
National Center for Theoretical Sciences, Hsinchu
300, Taiwan
}

\begin{abstract}
We study finite quantum corrections for several well known neutrino mixing matrices  and 
find that it is hard to account for  the large value of $\theta_{13}^{}$   recently reported by T2K and MINOS.
To nicely reproduce all experimentally favored neutrino mixing angles and masses,
we propose  a new neutrino mixing pattern.
We also demonstrate 
a simple realization by slightly extending the standard model to illustrate the   quantum corrections.
\end{abstract}

\maketitle

\section{Introduction}
During the recent decade, neutrino oscillation experiments have provided us with valuable information on the neutrino mixing angles and masses.
Despite the great progress, 
some of quantities in the neutrino sector have not been measured yet. 
In particular, one of the mixing angles,
$\theta_{13}^{}$,  has  been shrouded in mystery for a long time.
This angle has been believed to be very tiny compared with the other angles, and it has long been an open question whether  
its value is exactly zero or not. 
Over the years, people  have attempted to indirectly extract the hint of a non-zero $\theta_{13}$ via the precision analysis of oscillation channels 
with a gradually improved confidence level \cite{Fogli-fit, STV-fit, GMS-fit}. 
Recently, the T2K \cite{t2k} and MINOS \cite{MINOS} Collaborations have announced that they have observed some electron-like events in the $\nu_\mu \rightarrow \nu_e$ appearance channel, which is mainly sensitive to $\theta_{13}$, with the indications of a relatively large $\theta_{13}$: e.g., the best-fit value of $\theta_{13}$ suggested by the T2K Collaboration is around $9.7^\circ$ ($11.0^\circ$) for the normal (inverted) neutrino mass ordering.
Moreover, at almost the same time, the possible $>3\sigma$ evidence of a non-zero $\theta_{13}$ has been reported in a global analysis \cite{new-fit,new-STV}.

In reaction to the above developments, many theoretical studies have also been performed.
On one hand, in Ref. \cite{V0-del}, the authors focus on  mixing matrices which predict the vanishing $\theta_{13}$ at leading order and perturb it by introducing small corrections, so that the resultant $\theta_{13}$ can account for the T2K and MINOS indications.
On the other hand, new mixing patterns with a non-zero $\theta_{13}$ at leading order are also proposed in Ref. \cite{new-V0}\footnote{
Other studies relevant to the T2K and MINOS indications can be found in Ref. \cite{others}. 
For early works on a large $\theta_{13}^{}$, see Ref. \cite{early}.
}.
In this paper, by adopting the former paradigm, we also aim at deriving a large $\theta_{13}$ in a model-dependent manner.
In Ref. \cite{AGX},
 a new theoretical scheme was proposed, in which small corrections to the neutrino mixing angles and masses are induced from finite quantum effects. 
In this study, we introduce relatively large corrections to investigate whether $\theta_{13}$ can largely deviate from $0^\circ$ for the 
well known popular neutrino mixing patterns, such as
tri-bimaximal (TBM) \cite{TBM}, bi-maximal (BM) \cite{BM}, and democratic (DC) \cite{DC} mixing matrices.
Furthermore, we illustrate  a new (tree-level) mixing pattern, which can nicely reproduce all the experimental results after taking the finite quantum corrections into account.

This paper is organized as follows.
In Sec. II, we show a basic framework of our scheme and derive analytic expressions of the neutrino mixing angles and masses. 
In Sec. III, we numerically analyze the scheme for the TBM, BM, and DC mixing patterns.
We also demonstrate a new mixing pattern, which works very well with our scheme. 
In Sec. IV, we show a simple realization of the finite quantum corrections by extending the standard model (SM) with new $SU(2)_L^{}$ doublet and triplet scalars. 
We summarize the discussions in Sec. V.

\section{Finite quantum corrections}
We suppose that neutrinos are Majorana particles and divide the Majorana neutrino mass matrix ($M_\nu$)
into the tree-level ($M^0_\nu$) and one-loop ($\Delta M_\nu^{}$) parts 
\begin{eqnarray}
M_\nu^{} = M^0_\nu + \Delta M_\nu^{}. \label{eq:Mn}
\end{eqnarray}
The tree-level mass matrix is assumed to take the form of $M^0_\nu =  V^0_{} {\cal D}_\nu^{} (V^0_{})_{}^T$, where ${\cal D}_\nu^{} = \diag(\lambda_1^{}e^{i\rho}_{},\lambda_2^{}e^{i\sigma}_{},\lambda_3^{})$ contains the three eigenvalues of $M^0_\nu$ with two Majorana CP-violating phases, and $V^0_{}$ stands for the tree-level mixing matrix, so that $M^0_\nu$ is diagonalized by $V^0_{}$.
We also presume the diagonal charged lepton mass matrix and the existence of a flavor symmetry which ensures $\theta_{13}^{}=0^\circ$ at tree level.
Thus, we parametrize $V^0_{}$ as 
\begin{eqnarray}
V^0_{}=
\bmx{ccr}
 c_{12}^0 & s_{12}^0 & 0 \\
-c_{23}^0 s_{12}^0 & c_{23}^0 c_{12}^0 & -s_{23}^0 \\
-s_{23}^0 s_{12}^0 & s_{23}^0 c_{12}^0 &  c_{23}^0
\emx ,\label{eq:V0}
\end{eqnarray}
where $s_{ij}^0(c_{ij}^0)=\sin\theta_{ij}^0(\cos\theta_{ij}^0)$ with $\theta_{ij}^0$ representing the tree-level mixing angles. 
For this setup, we take
the following correction term: 
\begin{eqnarray}
\Delta M_\nu^{} = \frac{M_\nu^0 D_\ell^2 + D_\ell^2 M_\nu^0}{v^2_{}} \times I^{\rm loop}_{} , \label{eq:dM}
\end{eqnarray}
where $D_\ell^{}=\diag(m_e^{},m_\mu^{},m_\tau^{})$ denotes the diagonal charged lepton mass matrix, and $v=174~\gev$ is the vacuum-expectation-value (VEV) of the SM Higgs field.
Such a correction term may emerge from a one-loop diagram. We will show a simple realization of $\Delta M_\nu^{}$ in Sec. IV. 
Notice that $I^{\rm loop}_{}$ is a dimension-less function including the one-loop integral. 
Remarkably, in this scheme, once a specific $V^0_{}$ is given at tree level, the structure of $\Delta M_\nu^{}$ will be 
determined up to the over-all factor $I^{\rm loop}_{}$ because no new Yukawa coupling is introduced.

In principle, one also needs to consider the higher-loop (finite quantum) corrections to the neutrino mixing angles and masses.
However, to simplify our discussion on the generic feature of the corrections,
in what follows, we will assume that these contributions are negligible.
In fact, in the model presented in Sec. IV, the two-loop corrections are indeed always smaller than the one-loop correction by a factor $1/(16\pi^2_{})$.

We regard $\delta M_\nu$ as small perturbations, yielding the perturbed mixing angles: 
\begin{eqnarray}
\sin\theta_{13}^{}
&\simeq & 
\left| 2 s_{23}^0 c_{23}^0 s_{12}^0 c_{12}^0 \frac{m_\tau^2}{v^2_{}}
\left\{
\frac{\lambda_3^2[\lambda_1^2 + \lambda_2^2 - 2\lambda_1^{}\lambda_2^{}\cos(\rho - \sigma)]}{[\lambda_1^2 + \lambda_3^2 - 2\lambda_1^{}\lambda_3^{}\cos\rho][\lambda_2^2 + \lambda_3^2 - 2\lambda_2^{}\lambda_3^{}\cos\sigma]}
\right\}^{\frac{1}{2}}_{} I^{\rm loop}_{}\right|,
\\
\tan\theta_{12}
&\simeq&
t_{12}^0
\left[
1+(s_{23}^0)^2_{}\frac{m_\tau^2}{v^2_{}}\frac{\lambda_1^2 - \lambda_2^2}{\lambda_1^2 + \lambda_2^2 - 2\lambda_1^{}\lambda_2^{}\cos(\rho - \sigma)}I^{\rm loop}_{}
\right], \label{eq:psol}
\\
\tan\theta_{23}^{}
&\simeq&
t_{23}^0
\left\{ 
1+\frac{m_\tau^2}{v^2_{}}
\left[
\frac{(s_{12}^0)^2_{}(\lambda_1^2 - \lambda_3^2)}{\lambda_1^2 + \lambda_3^2 -2\lambda_1^{}\lambda_3^{}\cos\rho} + \frac{(c_{12}^0)^2_{}(\lambda_2^2 - \lambda_3^2)}{\lambda_2^2 + \lambda_3^2 -2\lambda_2^{}\lambda_3^{}\cos\sigma}
\right] I^{\rm loop}_{}
\right\}, \label{eq:patm}
\end{eqnarray}
the perturbed neutrino masses: 
\begin{eqnarray}
&&m_1^{} 
= \lambda_1^{}
\left[ 
1 + 2\frac{m_\tau^2}{v^2_{}}(s_{23}^0)^2_{}(s_{12}^0)^2_{} I^{\rm loop}_{}
\right], \\
&&m_2^{} 
= \lambda_2^{}
\left[
1 + 2\frac{m_\tau^2}{v^2_{}}(s_{23}^0)^2_{}(c_{12}^0)^2_{} I^{\rm loop}_{}
\right], \\
&&m_3^{} 
=\lambda_3^{}
\left[
1 + 2\frac{m_\tau^2}{v^2_{}}(c_{23}^0)^2_{} I^{\rm loop}_{}
\right],
\end{eqnarray}
and the Jarlskog parameter \cite{jarl}: 
\begin{eqnarray}
J_{\rm CP}^{} 
&\simeq & 
2 (s_{23}^0 c_{23}^0 s_{12}^0 c_{12}^0)^2_{} \frac{m_\tau^2}{v^2_{}}
\left[
\frac{\lambda_2^{}\lambda_3^{}\sin\sigma}{\lambda_2^2 + \lambda_3^2 -2\lambda_2^{}\lambda_3^{}\cos\sigma} - \frac{\lambda_1^{}\lambda_3^{}\sin\rho}{\lambda_1^2 + \lambda_3^2 -2\lambda_1^{}\lambda_3^{}\cos\rho}
\right] I^{\rm loop}_{} ,
\end{eqnarray}
where we have ignored the muon and electron masses.
In contrast, there are no corrections to $\rho$ and $\sigma$, which are Majorana CP-violating phases defined below Eq. (\ref{eq:Mn}), up to this order.

Here, we would emphasize two important features of the perturbed mixing angles, which will be crucial when we discuss the results of numerical calculations in the next section. 
(i) As found in the studies of renormalization-group equations \cite{rge}, corrections to the mixing angles can be enhanced due to the degeneracy among three neutrino masses. 
Particularly, in view of $\Delta m_{21}^2 \ll \Delta m_{31}^2$ and $\lambda_i^{} \simeq m_i^{}$, we can conjecture that $\theta_{12}^{}$ is the most sensitive one to this enhancement \cite{AGX}. 
However, because the strength of the enhancement depends also on the Majorana CP-violating phases as well as the differences between $\lambda_i^{}$ and $m_i^{}$, the enhancement is not always strong even in the case of the quasi-degenerate neutrino mass spectrum.
(ii) Relative signs between the tree-level mixing angles and their corrections can approximately be determined by the sign of $I^{\rm loop}_{}$ and the neutrino mass ordering in the sense of $\lambda_i^{}\simeq m_i^{}$. 
In fact, from Eqs. (\ref{eq:psol}) and (\ref{eq:patm}), one can immediately read out the following behaviors:
\begin{itemize}
\item 
$I^{\rm loop}_{}>(<)0~~$ leads to ~~$\theta_{12}^{}<(>)\theta_{12}^0$,
\item
NO with $I^{\rm loop}_{}>(<)0$ and IO with $I^{\rm loop}_{}<(>)0$~~ yield ~~$\theta_{23}^{}<(>)\theta_{23}^0$,
\end{itemize}
where NO (IO) denotes the normal (inverted) neutrino mass ordering.

\section{Numerical calculations}
\subsection{Input parameters}
Instead of a perturbative method, we numerically diagonalize the full neutrino mass matrix of Eq. (\ref{eq:Mn}) and compute the neutrino mixing angles and masses.
 From the recent global analysis \cite{STV-fit} of the neutrino oscillation data, we refer to the following best-fit values and $1\sigma$ ($3\sigma$) error bounds: 
\begin{eqnarray}
&&\Delta m_{21}^2 = 
\left(7.59^{+0.20(0.60)}_{-0.18(0.50)}\right)
\times 10^{-5}_{}~~\ev^2 ,
\nonumber \\
&&\Delta m_{31}^2 =
\left\{\begin{array}{ll}
+\left(2.45_{-0.09(0.27)}^{+0.09(0.28)}\right)
\times 10^{-3}_{}~~\ev^2 & ~~{\rm for~Normal~Ordering~(NO)} \\
-\left(2.34^{+0.10(0.30)}_{-0.09(0.26)}\right) 
\times 10^{-3}_{}~~\ev^2 & ~~{\rm for~Inverted~Ordering~(IO)}
\end{array}\right. ,\nonumber \\
&&\theta_{12}^{} = 
\left(34.0^{+1.0(2.9)}_{-1.0(2.7)}\right)^\circ,
~~~~
\theta_{23}^{} = 
\left\{\begin{array}{l}
\left(45.6^{+3.4(7.5)}_{-3.5(7.0)}\right)^\circ \\ 
\left(46.1^{+3.5(7.0)}_{-3.4(7.5)}\right)^\circ
\end{array}\right. ,
~~~~
\theta_{13}^{} = 
\left\{\begin{array}{l}
\left(5.7^{+2.2(5.1)}_{-2.1(--)}\right)^\circ \\
\left(6.5^{+2.0(4.9)}_{-2.1(--)}\right)^\circ
\end{array}\right. , \label{eq:gfit}
\end{eqnarray}
where the upper and lower values of $\Delta m_{31}^2$, $\theta_{23}$, and $\theta_{13}$ correspond to the NO and IO, respectively.
In the following calculations, unless otherwise stated, we impose the $3\sigma$ constraints on $\Delta m_{21}^2$, $\Delta m_{31}^2$, and $\theta_{12}^{}$ to examine $\theta_{13}^{}$ and $\theta_{23}^{}$ as well as $J_{\rm CP}^{}$. 
Besides,   we use the charged lepton masses at the electroweak scale as~\cite{xzz}
\begin{eqnarray}
m_e^{}=0.486~\mev,~~
m_\mu^{}=102.718~\mev,~~
m_\tau^{}=1746.24~\mev .
\end{eqnarray}

We vary 
the Majorana phases ($\rho$ and $\sigma$) within 
$0^\circ$ to $360^\circ$
and $\lambda_i^{}$ to fit the two mass-squared differences.
Moreover, in order to enhance the corrections to the mixing angles, we consider the quasi-degenerate neutrino mass spectrum and fix the heaviest neutrino mass as $0.2~\ev$, i.e., $m^{}_{3(2)}=0.2~\ev$ in the case of the NO (IO).
In this case, $|I^{\rm loop}_{}|\simeq 100$ is needed to realize $\theta_{13}^{}\simeq 10^\circ$.
Here, we vary $I^{\rm loop}_{}$ within $-125$ to $125$, so that 
a maximal value of $\theta_{13}^{}\simeq 13^\circ$ can be produced.
We note that almost the same results can be obtained for the different choices of $m_{3(2)}$ and $I^{\rm loop}_{}$, e.g., $m_{3(2)}=0.3~\ev$ with $|I^{\rm loop}_{}| < 55$ or  $m_{3(2)}=0.4~\ev$ with $|I^{\rm loop}_{}| < 30$.

\subsection{Tri-bimaximal (TBM) mixing}
\begin{figure}[ht]
\begin{center}
\includegraphics*[width=0.49\textwidth]{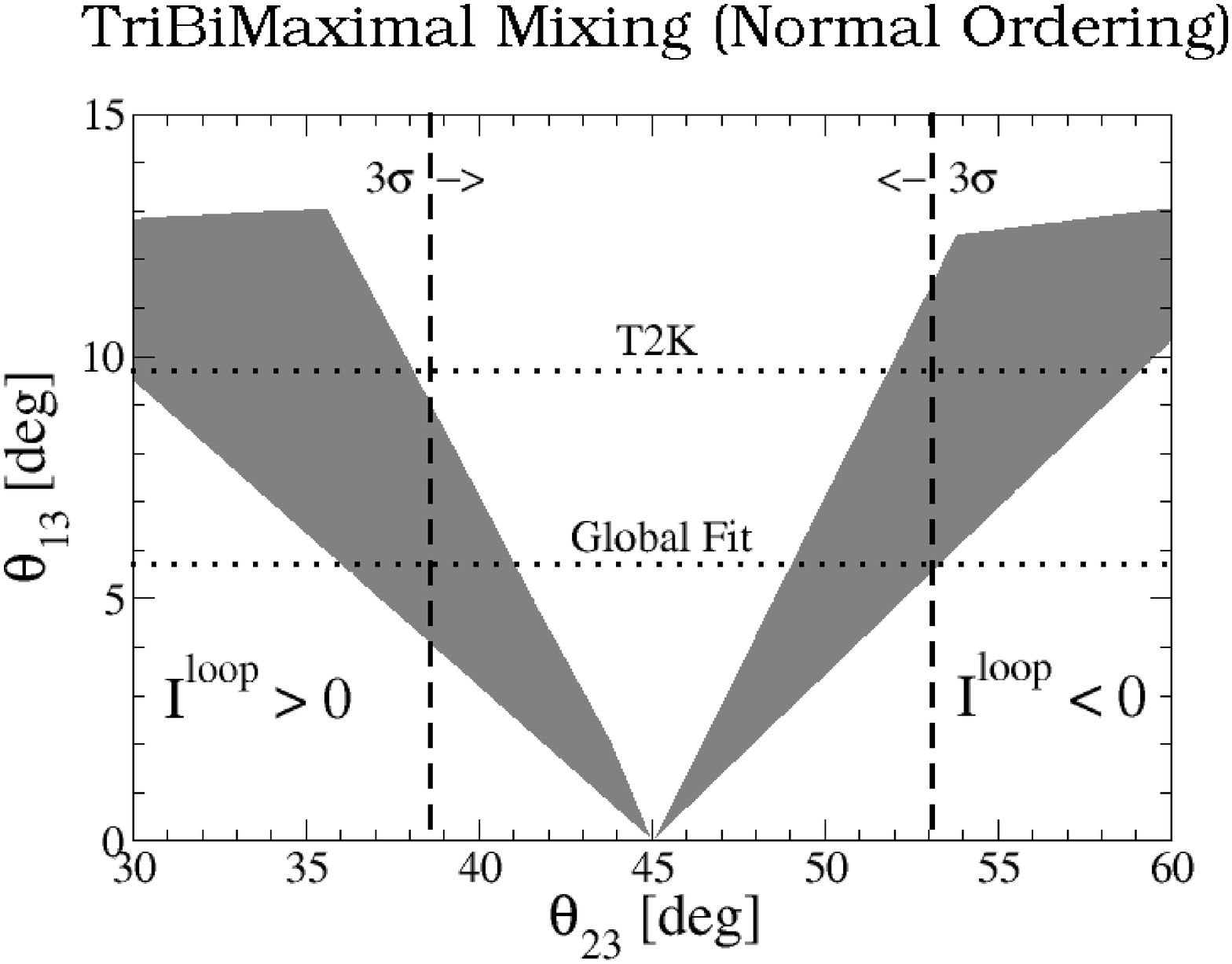}
\includegraphics*[width=0.49\textwidth]{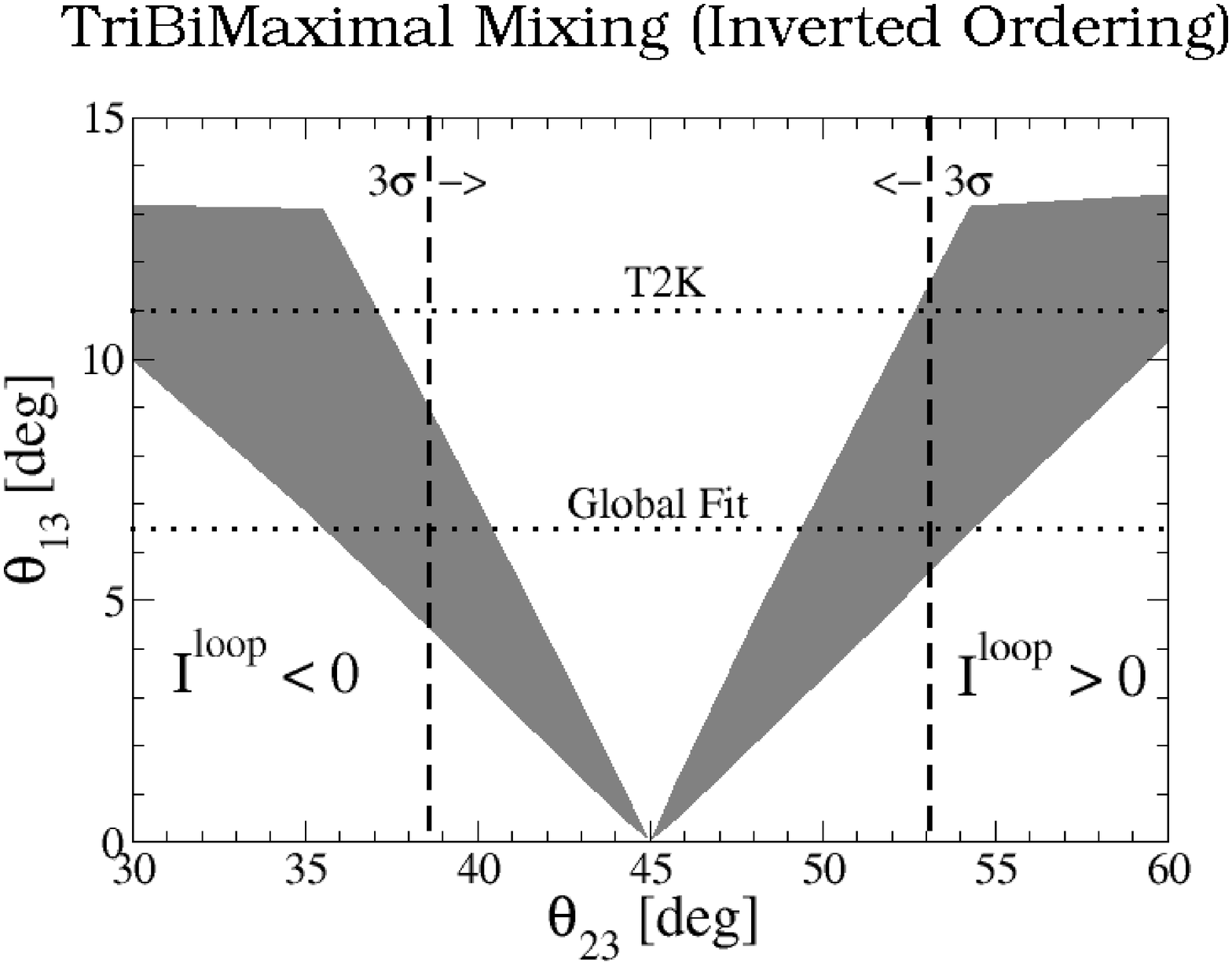}
\vspace{-0.3cm}
\caption{\footnotesize 
$\theta_{13}^{}$ as functions of $\theta_{23}^{}$ for the normal ordering (NO)  and inverted ordering (IO) at the left and  right panels, respectively, with
  the tri-bimaximal (TBM) tree-level mixing matrix, 
  where the horizontal doted lines display the best-fit values of $\theta_{13}^{}$ from the T2K experiment \cite{t2k} and Eq. (\ref{eq:gfit}), while the vertical dashed lines express the $3\sigma$ upper and lower bounds of $\theta_{23}$ from Eq. (\ref{eq:gfit}).
} \label{fig:ra-TB}
\end{center}
\end{figure}
\begin{figure}[h]
\begin{center}
\includegraphics*[width=0.49\textwidth]{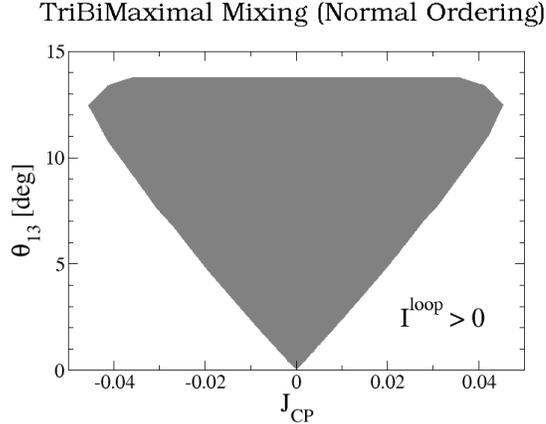}
\vspace{-0.3cm}
\caption{\footnotesize 
$\theta_{13}^{}$ as a function of $J_{\rm CP}^{}$ for the normal ordering (NO) with $I^{\rm loop}_{}>0$ in the case where the tree-level mixing matrix is the tri-bimaximal (TBM) one.
} \label{fig:rj-TB}
\end{center}
\end{figure}
Let us employ the TBM mixing 
\begin{eqnarray}
V^0_{\rm TB}=\frac{1}{\sqrt{6}}
\bmx{ccr}
 2 & \sqrt{2} & 0 \\
-1 & \sqrt{2} & -\sqrt{3} \\
-1 & \sqrt{2} &  \sqrt{3}
\emx 
\end{eqnarray}
as the tree-level mixing matrix. Namely, we substitute $\theta_{12}^0 \simeq 35.26^\circ$ and $\theta_{23}^0=45^\circ$ in Eq. (\ref{eq:V0}).
In Fig. \ref{fig:ra-TB}, we compute $\theta_{13}^{}$ as functions of $\theta_{23}^{}$ for the NO (left panel) and IO (right panel), respectively.
As can be seen, $\theta_{13}^{}$ can largely deviate from $0^\circ_{}$ and it can even be $10^\circ \sim 11^\circ$,  favored by the T2K experiment \cite{t2k}.
However, such a large $\theta_{13}$ simultaneously leads to a large deviation of $\theta_{23}^{}$ from $45^\circ_{}$.
For instance, if we demand the $3\sigma$ constraint of $\theta_{23}^{}$ in Eq. (\ref{eq:gfit}), $\theta_{13}^{}$ can maximally be $9.0^\circ$ ($11.5^\circ$) and $11.2^\circ$ ($9.0^\circ$) for $I^{\rm loop}_{}>0$ and $I^{\rm loop}_{}<0$, respectively, in the case of the NO (IO), but these values correspond to the edges of $3\sigma$ upper and lower bounds of $\theta_{23}$. 
Therefore, $\theta_{13}^{}\simeq 10^\circ$ cannot be accompanied with a nearly maximal $\theta_{23}^{}$, which is favored by the neutrino oscillation data, in the TBM mixing case.
In Fig. \ref{fig:rj-TB}, we also plot $\theta_{13}^{}$ as a function of $J_{\rm CP}^{}$ for only the NO with $I^{\rm loop}_{}>0$ case and find that $J_{\rm CP}^{}$ can be of  ${\cal O}(0.01)$.
In this plane, the sign of $I^{\rm loop}_{}$ and the neutrino mass ordering do not make a large difference to the shape of allowed regions. 
Note that the $I^{\rm loop}<0$ case is excluded at the $1\sigma$ level due to $\theta_{12}^{}>35.26^\circ$,  mentioned at the end of Sec. II.

\subsection{Bi-maximal (BM) mixing}
\begin{figure}[ht]
\begin{center}
\includegraphics*[width=0.49\textwidth]{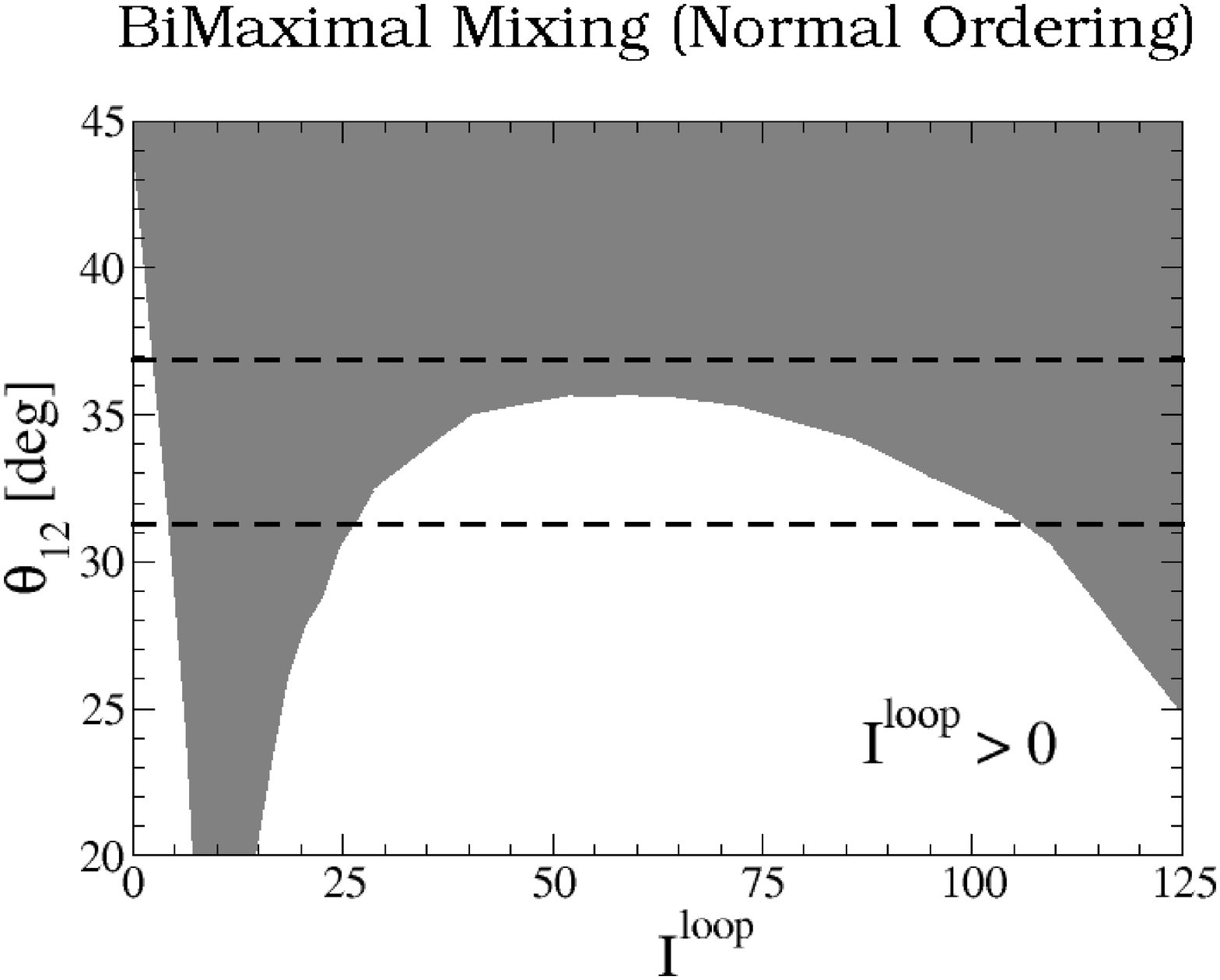}
\includegraphics*[width=0.49\textwidth]{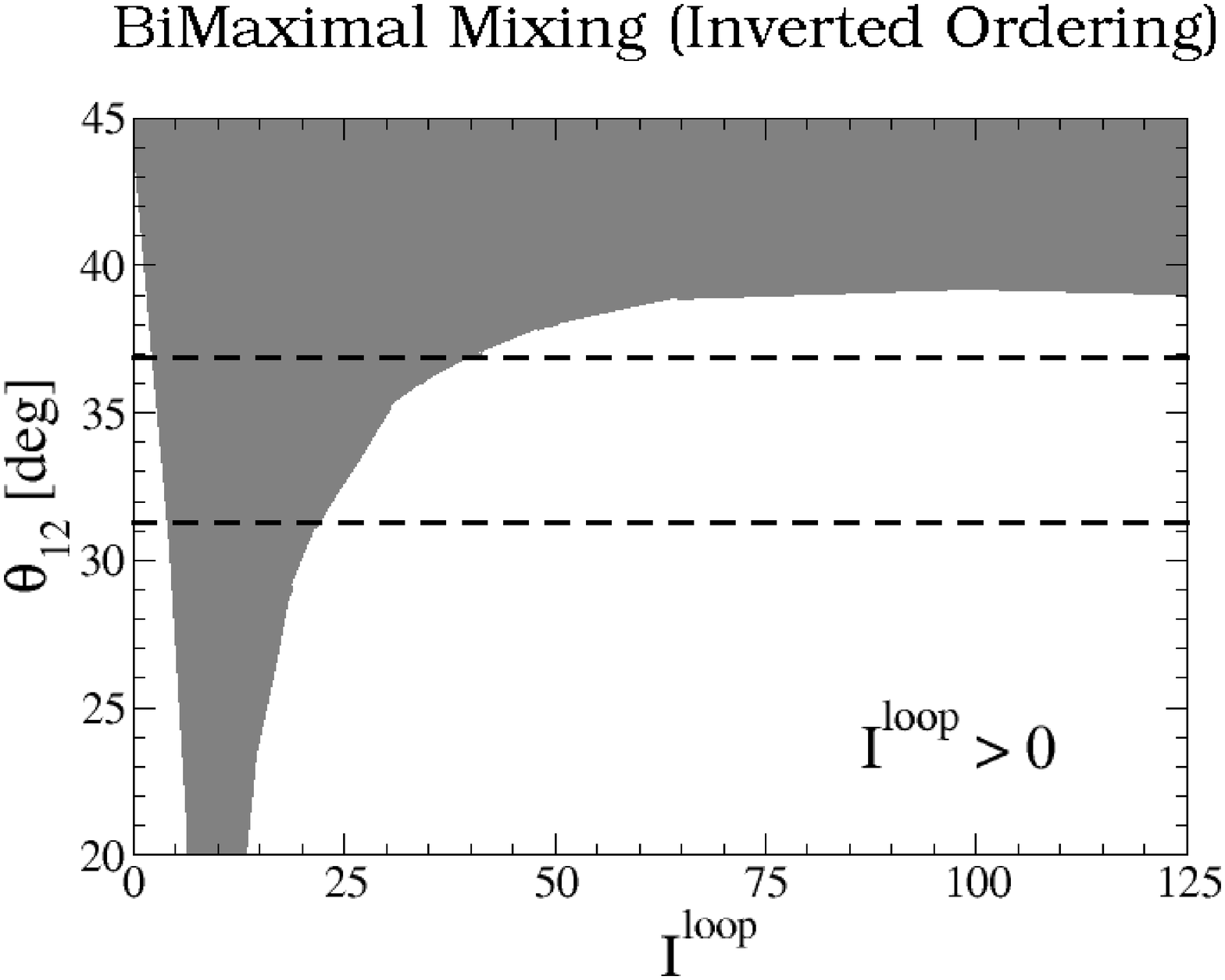}
\vspace{-0.3cm}
\caption{\footnotesize 
$\theta_{12}^{}$ as functions of $I^{\rm loop}_{}>0$
for the normal ordering (NO)  and inverted ordering (IO) at the left and right panels, respectively,  with 
  the bi-maximal (BM) tree-level mixing matrix, 
where the dashed lines display the $3\sigma$ upper and lower bounds of $\theta_{12}^{}$ from Eq. (\ref{eq:gfit}).
} \label{fig:si-BM}
\end{center}
\end{figure}
\begin{figure}[ht]
\begin{center}
\includegraphics*[width=0.49\textwidth]{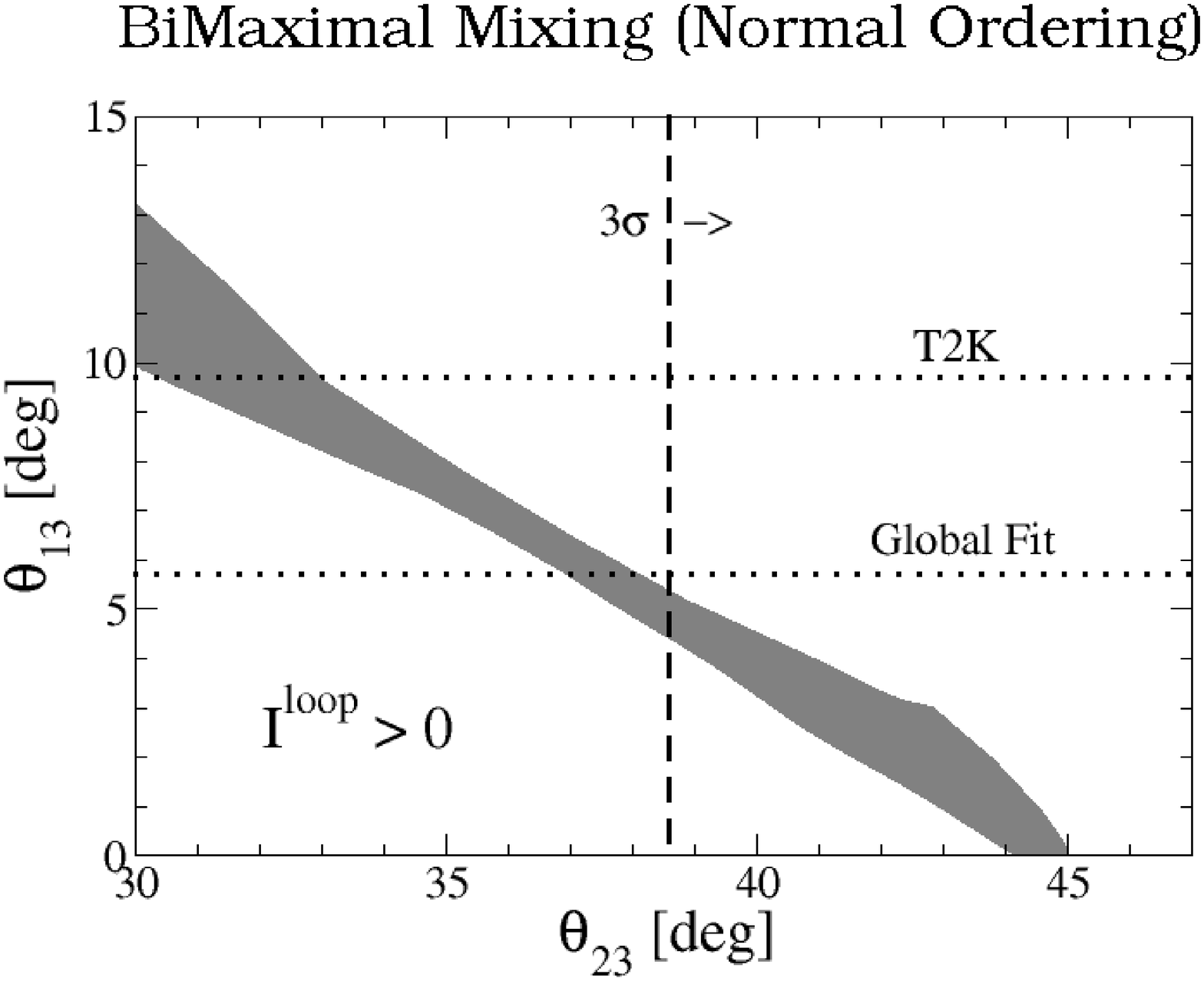}
\includegraphics*[width=0.49\textwidth]{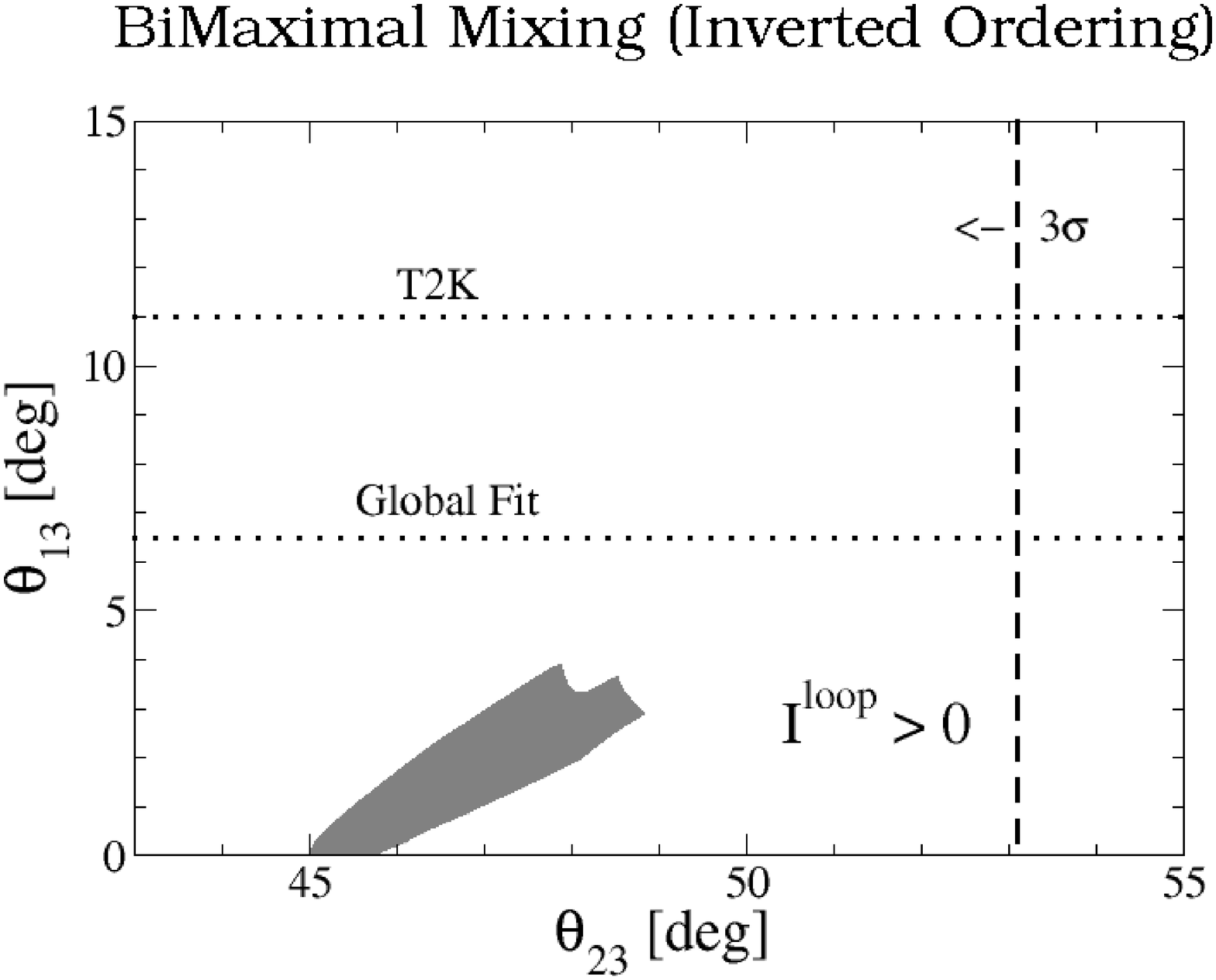}
\vspace{-0.3cm}
\caption{\footnotesize 
Legend is the same as Fig. \ref{fig:ra-TB} except that the tree-level mixing matrix is the bi-maximal (BM)  one with $I^{\rm loop}_{}>0$.
} \label{fig:ra-BM}
\end{center}
\end{figure}
\begin{figure}[ht]
\begin{center}
\includegraphics*[width=0.49\textwidth]{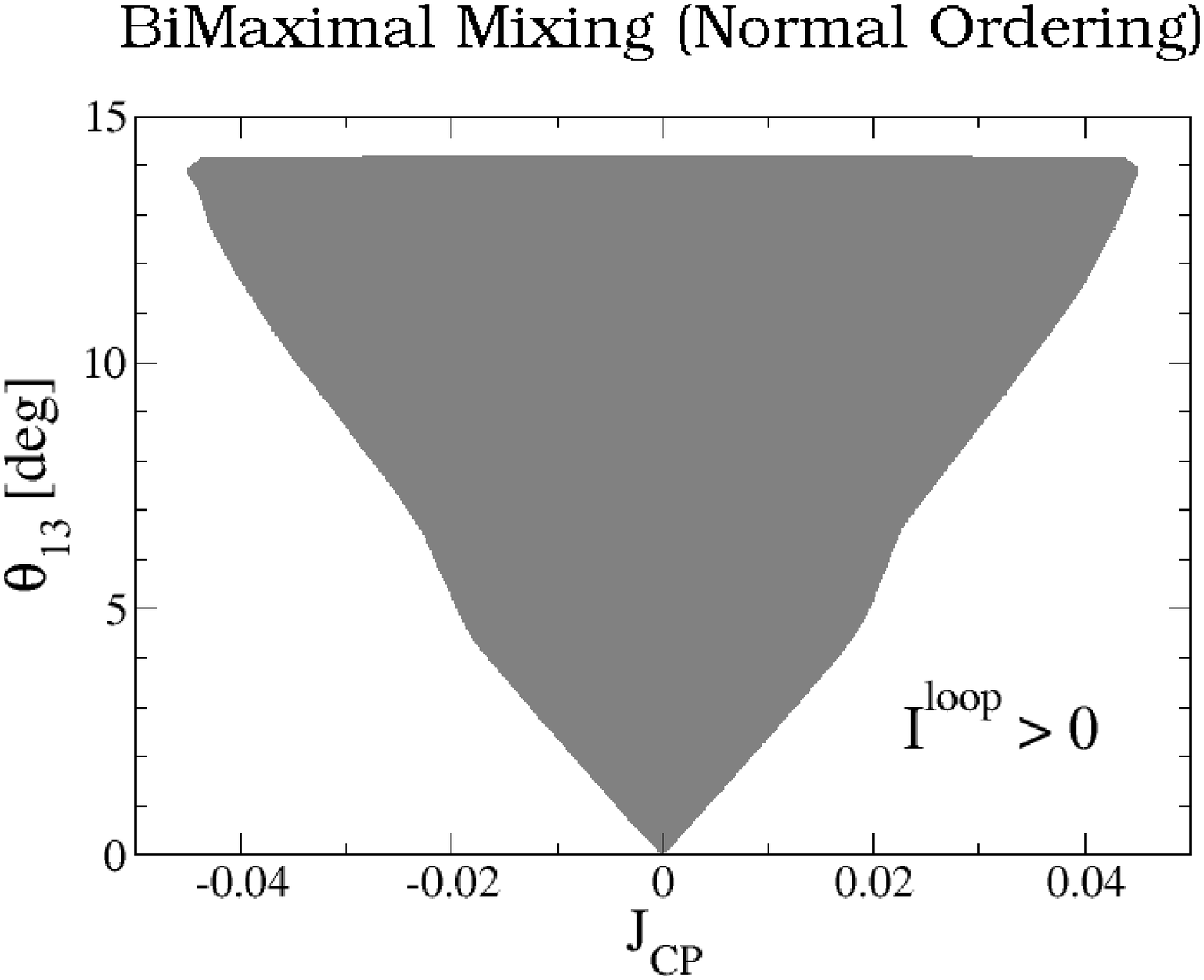}
\includegraphics*[width=0.49\textwidth]{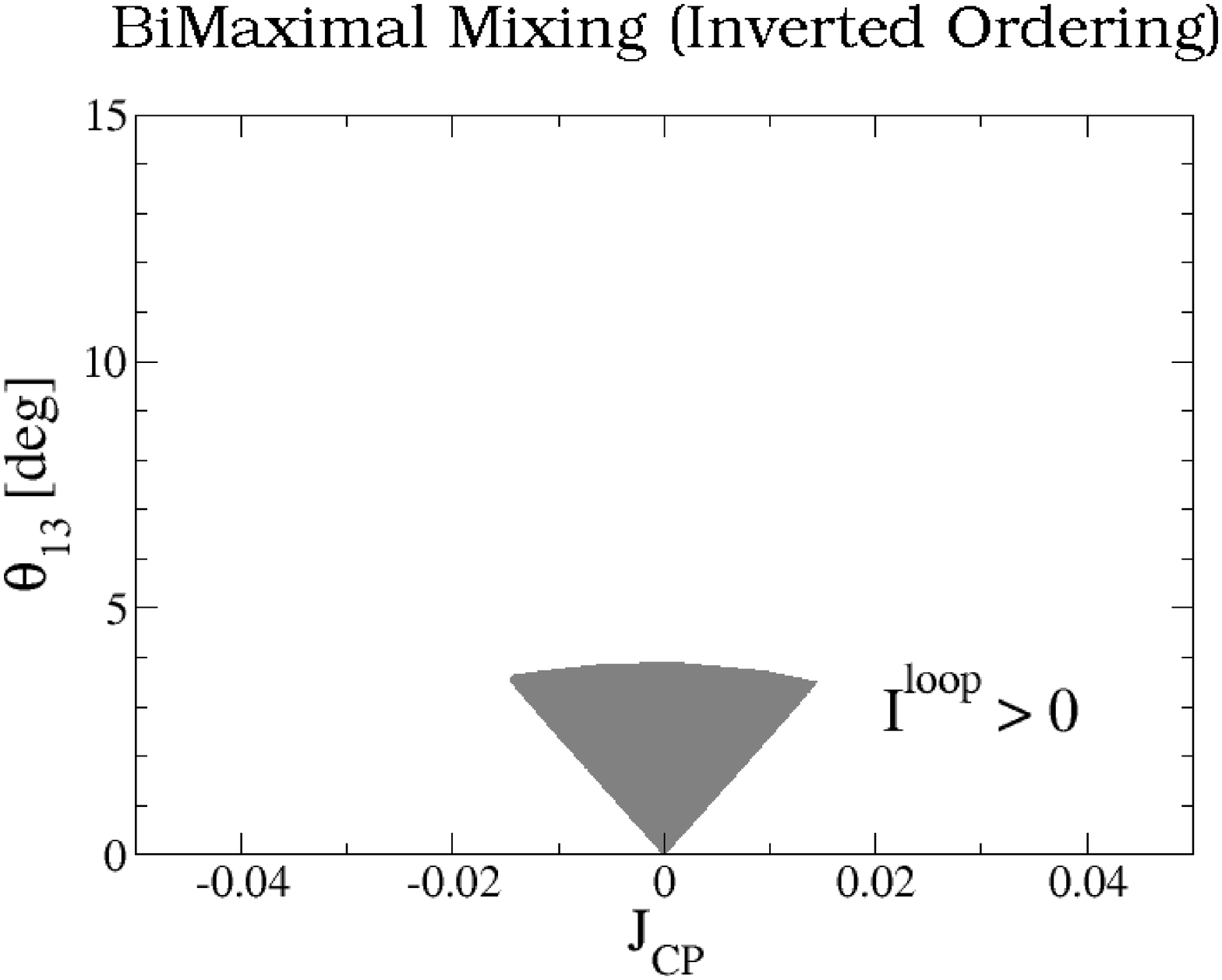}
\vspace{-0.3cm}
\caption{\footnotesize 
$\theta_{13}^{}$ as functions of $J_{\rm CP}^{}$ for the normal ordering (NO)  and inverted ordering (IO) at the left and  right panels, respectively, with $I^{\rm loop}_{}>0$ in the case where the tree-level mixing matrix is the bi-maximal (BM)  one.
} \label{fig:rj-BM}
\end{center}
\end{figure}
We use the BM mixing
\begin{eqnarray}
V^0_{\rm BM}=\frac{1}{2}
\bmx{ccr}
 \sqrt{2} & \sqrt{2} & 0 \\
-1 & 1 & -\sqrt{2} \\
-1 & 1 &  \sqrt{2}
\emx 
\end{eqnarray}
as the tree-level mixing matrix with $\theta_{12}^0 = \theta_{23}^0=45^\circ$ for Eq. (\ref{eq:V0}). 
In the case of $I^{\rm loop}_{}<0$, this tree-level mixing results in $\theta_{12}^{}>45^\circ$, and it is clearly inconsistent with experiments. 
Furthermore, even in the case of $I^{\rm loop}>0$, $\theta_{12}$ cannot always account for the $3\sigma$ upper bound ($\theta_{12}<36.9^\circ$), so that the allowed regions are restricted in comparison with those of the TBM case. 
To illustrate the behavior of $\theta_{12}^{}$, in Fig. \ref{fig:si-BM}, we calculate $\theta_{12}^{}$ as a function of $I^{\rm loop}_{}$ with respect to only the $3\sigma$ constraints of $\Delta m_{21}^2$ and $\Delta m_{31}^2$.
As one can see, the corrections of $\theta_{12}^{}$ are maximally enhanced around $I^{\rm loop}_{} \simeq 10$, but the enhancement becomes weaker as $I^{\rm loop}_{}$ increases.
Especially, in the IO case, $\theta_{12}$ gets always away from the $3\sigma$ range after $I^{\rm loop}_{} \simeq 40$. 
In turn, we restore the $3\sigma$ constraint of $\theta_{12}^{}$ and plot $\theta_{13}^{}$ as functions of $\theta_{23}^{}$ and $J_{\rm CP}^{}$ in Figs. \ref{fig:ra-BM} and \ref{fig:rj-BM}, respectively.
In the case of the NO, $\theta_{13}$ can maximally be $5.3^\circ$ at the $3\sigma$ lower bound of $\theta_{23}$, while it can be $4.0^\circ$ at $\theta_{23}\simeq 48^\circ$ (and $\theta_{12}\simeq 36.9^\circ$) in the IO case. 
Hence, we conclude that the BM mixing case cannot explain $\theta_{13}^{}\simeq 10^\circ$, favored by the T2K experiment, while keeping the other angles within experimentally favored ranges.
Also, the allowed region of $J_{\rm CP}^{}$ is strictly limited in the IO case.

\subsection{Democratic (DC) mixing}
\begin{figure}[ht]
\begin{center}
\includegraphics*[width=0.49\textwidth]{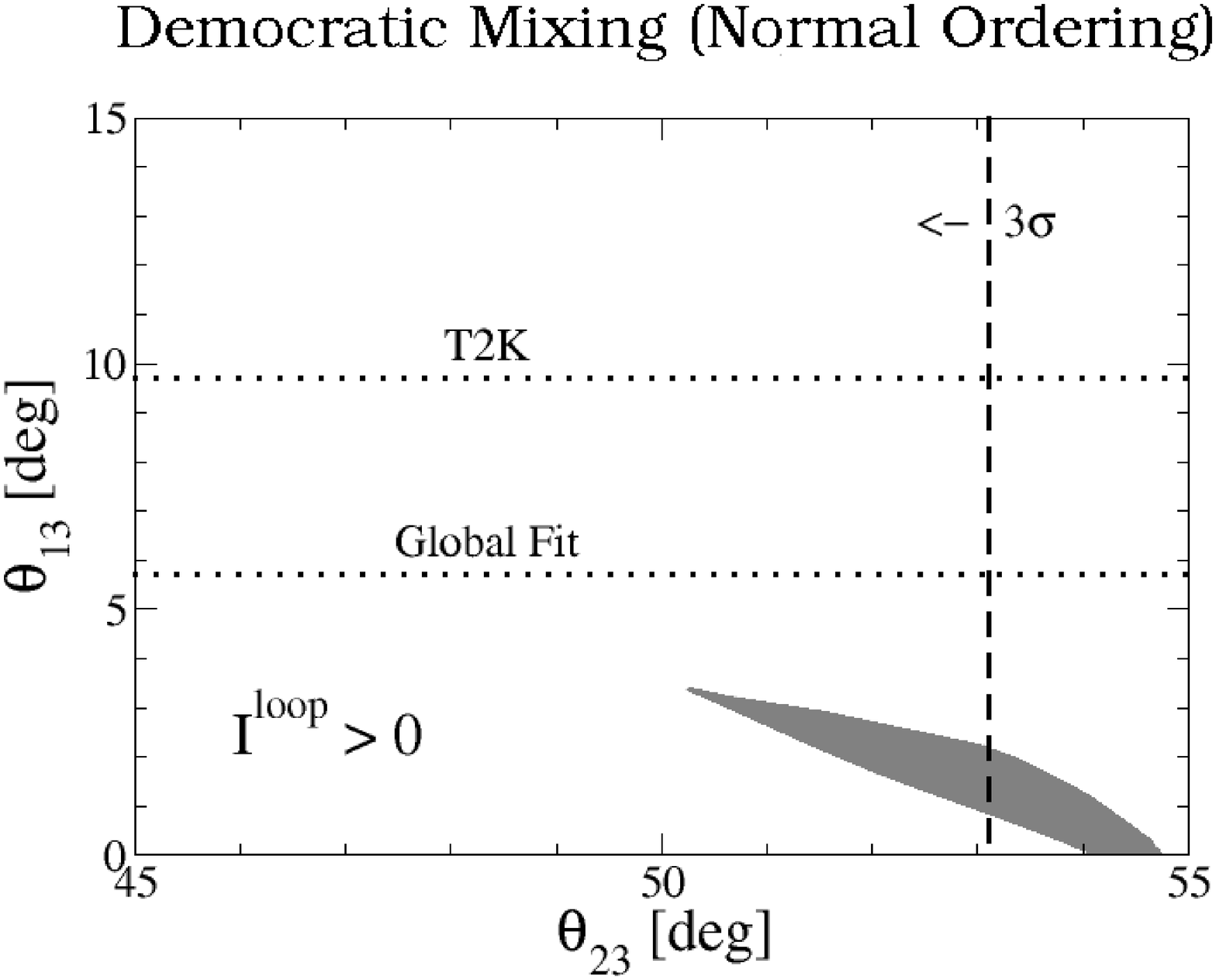}
\includegraphics*[width=0.49\textwidth]{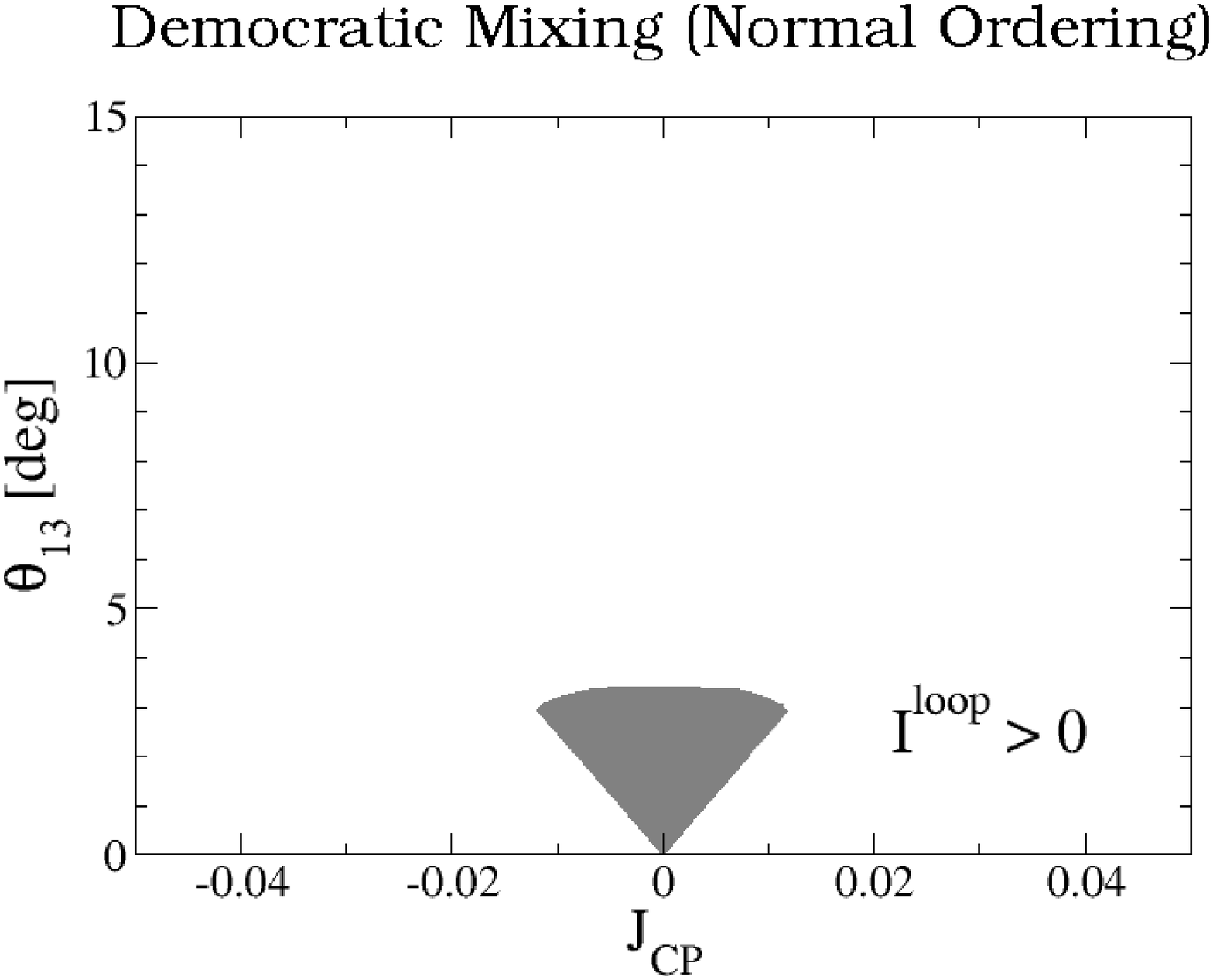}
\vspace{-0.3cm}
\caption{\footnotesize 
Legend is  the same as Figs. \ref{fig:ra-TB} and \ref{fig:rj-TB} but with 
the democratic (DC) tree-level mixing matrix and $I^{\rm loop}_{}>0$.
} \label{fig:DC}
\end{center}
\end{figure}
We take the DC mixing
\begin{eqnarray}
V^0_{\rm DC}=\frac{1}{\sqrt{6}}
\bmx{ccr}
 \sqrt{3} & \sqrt{3} & 0 \\
-1 & 1 & -2 \\
-\sqrt{2} & \sqrt{2} &  \sqrt{2}
\emx 
\end{eqnarray}
as the tree-level mixing matrix with $\theta_{12}^0 = 45^\circ$ and $\theta_{23}^0 \simeq 54.74^\circ$ for Eq. (\ref{eq:V0}).
In the case of $I^{\rm loop}_{}<0$, this tree-level mixing works out $\theta_{12}^{}>45^\circ$, and it is clearly inconsistent with experiments.
Similarly, the IO with $I^{\rm loop}_{}>0$ case results in $\theta_{23}^{}>54.74^\circ$, and this case is obviously disfavored by experiments, too. 
Thus, the only possible combination is the NO with $I_{}^{\rm loop}>0$ one.
Even in this case, however, the allowed regions are strictly constrained by $\theta_{12}^{}$, like the BM case. 
 From Fig. \ref{fig:DC}, one can read off that
the maximum deviation of $\theta_{13}^{}$ from $0^\circ$ is only $3.4^\circ$ at $\theta_{23}^{}\simeq 50^\circ$ (and $\theta_{12}\simeq 36.9^\circ$), which indicates that the DC mixing is incompatible with our scheme to reproduce the large value of $\theta_{13}^{}$.

\subsection{New mixing}
\begin{figure}[ht]
\begin{center}
\includegraphics*[width=0.49\textwidth]{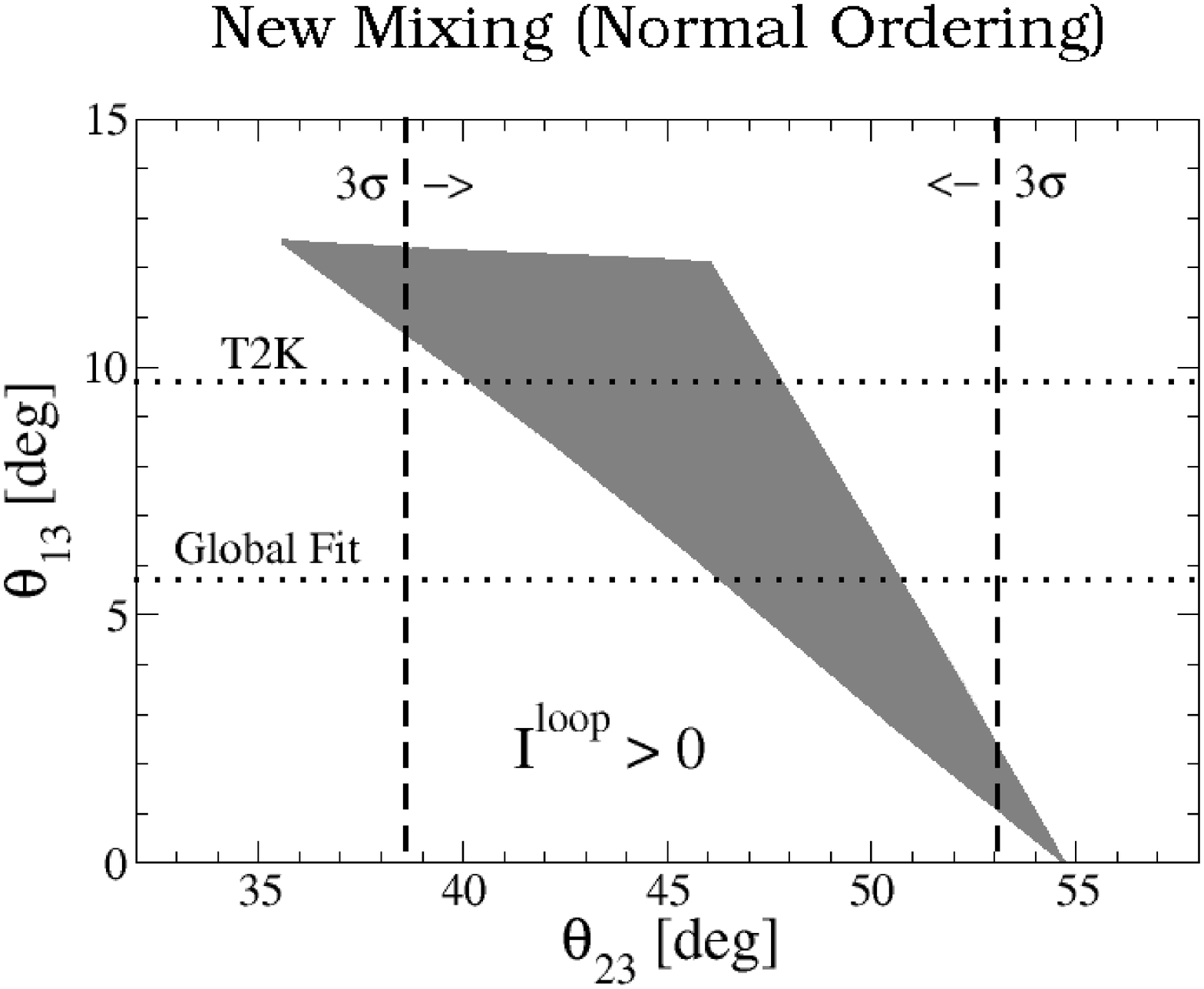}
\includegraphics*[width=0.49\textwidth]{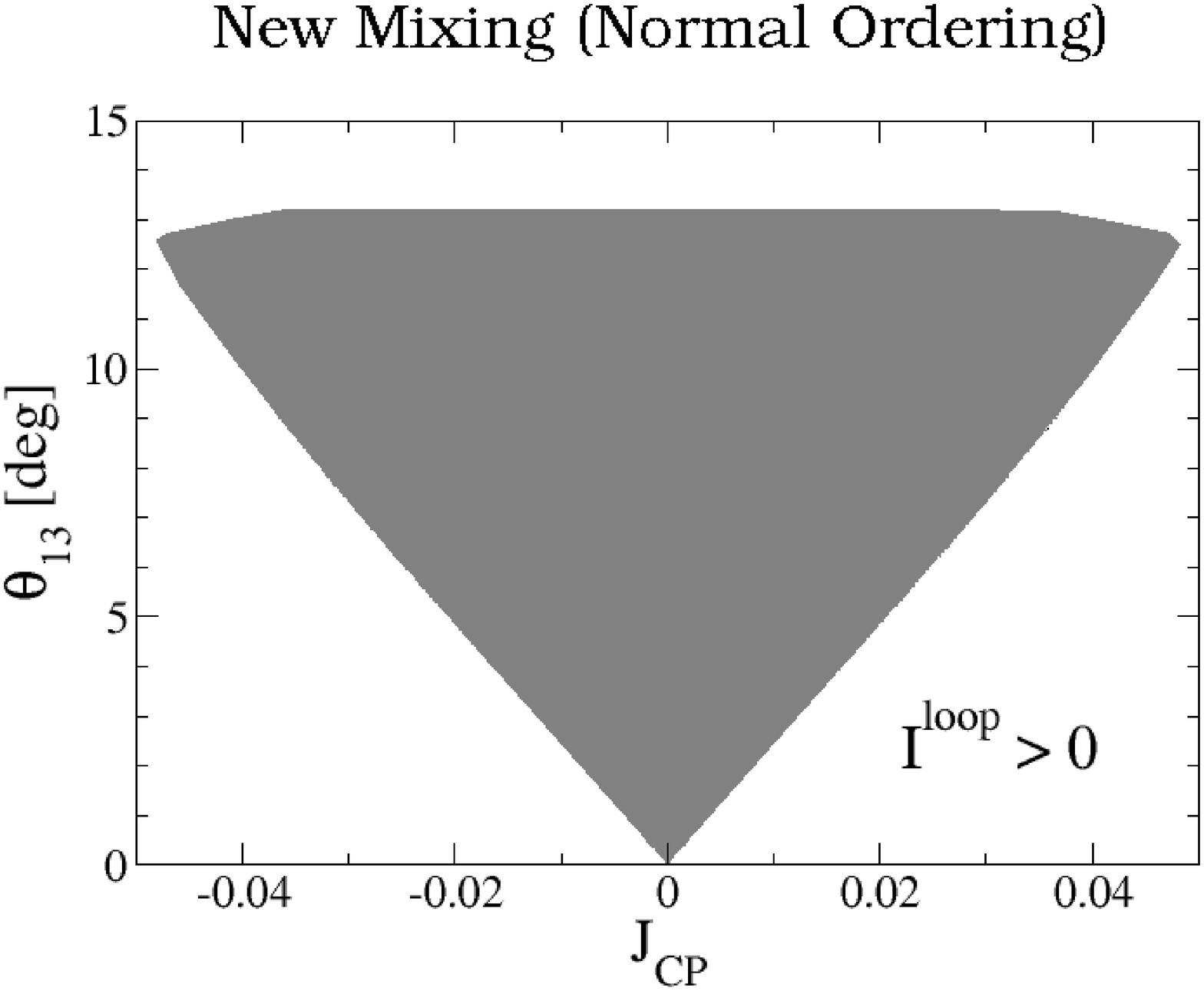}
\vspace{-0.3cm}
\caption{\footnotesize 
$\theta_{13}^{}$ as  functions of $\theta_{23}^{}$ (left panel) and $J_{\rm CP}^{}$ (right panel) for the normal ordering (NO) with $I^{\rm loop}_{}>0$ in the case where the tree-level mixing matrix takes the from of Eq. (\ref{eq:new}). 
Legends of the lines are the same as Fig.~\ref{fig:ra-TB}.
} \label{fig:new}
\end{center}
\end{figure}
From results we have obtained so far, we can read out several tendencies of our scheme: (i) $\theta_{13}$ can largely deviate from $0^\circ$ with a relatively large $I^{\rm loop}$, (ii) the large deviation of $\theta_{13}$ is always accompanied with a large deviation of $\theta_{23}$ from its initial value, and (iii) in the large $I^{\rm loop}$ region, $\theta_{12}$ cannot always  drastically depart from its initial value.
Due to these tendencies, we invent a new mixing pattern 
\begin{eqnarray}
V^0_{\rm new}=\frac{1}{3}
\bmx{ccr}
 \sqrt{6} & \sqrt{3} & 0 \\
-1 & \sqrt{2} & -\sqrt{6} \\
-\sqrt{2} & 2 &  \sqrt{3}
\emx\,, \label{eq:new}
\end{eqnarray}
 which predicts $\theta_{12}^0 = 35.26^\circ$,  $\theta_{23}^0=54.74^\circ$, and $\theta_{13}^0=0^\circ$ at tree level. 
We will demonstrate that this mixing matrix
can nicely reproduce all the experimental results after taking account of the finite quantum corrections.
However, for the NO with $I^{\rm loop}<0$ and IO with $I^{\rm loop}>0$, 
the mixing matrix in Eq. (\ref{eq:new})
results in $\theta_{23}>54.74^\circ$. 
Hence, we will concentrate on  the cases of the NO with $I^{\rm loop}>0$ and IO with $I^{\rm loop}<0$. 
In Fig. \ref{fig:new}, we plot $\theta_{13}^{}$ as functions of $\theta_{23}^{}$ (left panel) and $J_{\rm CP}^{}$ (right panel) for the NO one. 
Figures for the IO case are almost the same as Fig. \ref{fig:new}.
The  allowed regions in the figures are compatible with experiments very well. Consequently,  $\theta_{13}^{}$ can account for the best-fit values of the T2K experiment and Eq. (\ref{eq:gfit}) around the best-fit value of $\theta_{23}^{}$.
Moreover, we compute $\chi^2$ for $\Delta m_{21}^2$, $\Delta m_{31}^2$, $\theta_{12}^{}$, $\theta_{23}^{}$, and $\theta_{13}^{}$ based on the best-fit values and $1\sigma$ errors given in 
both Eq. (\ref{eq:gfit}) and
Ref. \cite{new-fit}:
\begin{eqnarray}
&&\Delta m_{21}^2 = \left(7.58^{+0.22}_{-0.26}\right)\times 10^{-5}_{}~~\ev^2 ,
~~~~
|\Delta m_{31}^2| = \left(2.35_{-0.09}^{+0.12}\right)\times 10^{-3}_{}~~\ev^2 ,\nonumber \\
&&\theta_{12}^{} = 
\left(34.0\pm 1.0\right)^\circ,
~~~~
\theta_{23}^{} = 
\left(40.4^{+4.6}_{-1.8}\right)^\circ ,
~~~~
\theta_{13}^{} = 
\left(9.1^{+1.2}_{-1.4}\right)^\circ , 
\label{eq:nfit}
\end{eqnarray}
in which new T2K and MINOS results are involved.
At the point where $\chi^2$ turns out to be minimum, based on Eqs. (\ref{eq:gfit}) and (\ref{eq:nfit}) we obtain the results as follows:
\begin{eqnarray}
&&\Delta m_{21}^2=7.59\times 10^{-5}~\ev^2, ~~
\Delta m_{31}^2=2.47\times 10^{-3}~\ev^2, \nonumber \\
&&\theta_{12}=34.1^\circ,~~
\theta_{23}=46.6^\circ,~~
\theta_{13}=6.3^\circ
\end{eqnarray}
and 
\begin{eqnarray}
&&\Delta m_{21}^2=7.61\times 10^{-5}~\ev^2, ~~
\Delta m_{31}^2=2.36\times 10^{-3}~\ev^2, \nonumber \\
&&\theta_{12}=33.8^\circ,~~
\theta_{23}=41.3^\circ,~~
\theta_{13}=9.1^\circ
\end{eqnarray}
for the NO, and
\begin{eqnarray}
&&\Delta m_{21}^2=7.63\times 10^{-5}~\ev^2, ~~
\Delta m_{31}^2=2.33\times 10^{-3}~\ev^2, \nonumber \\
&&\theta_{12}=34.7^\circ,~~
\theta_{23}=45.5^\circ,~~
\theta_{13}=6.8^\circ
\end{eqnarray}
and 
\begin{eqnarray}
&&\Delta m_{21}^2=7.62\times 10^{-5}~\ev^2,~~ 
\Delta m_{31}^2=2.35\times 10^{-3}~\ev^2, \nonumber \\
&&\theta_{12}=34.3^\circ,~~
\theta_{23}=41.2^\circ,~~
\theta_{13}=9.4^\circ
\end{eqnarray}
for the IO, respectively.

\subsection{Summary of numerical calculations}
\begin{table}[h]
\begin{center}
\begin{tabular}{|c|c|c||c|c|c|}\hline
Mixing & Ordering & $I^{\rm loop}$ & $\theta_{13}^{}$ & $\theta_{23}^{}$ & $J_{\rm CP}^{}$
\\ \hline\hline
  & NO & $+$ & $0.0^\circ \sim \underline{9.0^\circ}$ & $\underline{38.6^\circ} \sim 45.0^\circ$ & $0.0 \pm \underline{0.033}$
\\ \cline{3-6}
TBM &  & $-$ & ~~$0.0^\circ \sim \overline{11.2^\circ}$~~ & ~~$45.0^\circ \sim \overline{53.1^\circ}$~~ & ~~$0.0 \pm \overline{0.041}$~~
\\ \cline{2-6}
 & IO & $+$ & $0.0^\circ \sim \overline{11.5^\circ}$ & $45.0^\circ \sim \overline{53.1^\circ}$ & $0.0 \pm \overline{0.042}$
\\ \cline{3-6}
 &  & $-$ & $0.0^\circ \sim \underline{9.0^\circ}$ & $\underline{38.6^\circ} \sim 45.0^\circ$ & $0.0 \pm \underline{0.033}$
\\ \hline\hline
  & NO & $+$ & $0.0^\circ \sim \underline{5.3^\circ}$ & $\underline{38.6^\circ} \sim 45.0^\circ$ & $0.0 \pm \underline{0.018}$
\\ \cline{3-6}
BM &  & $-$ & \multicolumn{3}{|c|}{excluded}
\\ \cline{2-6}
 & IO & $+$ & $0.0^\circ \sim \langle3.9^\circ\rangle$ & $45.0^\circ \sim 48.9^\circ$ & $0.0 \pm 0.014$
\\ \cline{3-6}
 &  & $-$ & \multicolumn{3}{|c|}{excluded}
\\ \hline\hline
  & NO & $+$ & $\overline{0.9^\circ} \sim \langle3.4^\circ\rangle$ & $50.2^\circ \sim \overline{53.1^\circ}$ & $0.0 \pm 0.012$
\\ \cline{3-6}
DC &  & $-$ & \multicolumn{3}{|c|}{excluded}
\\ \cline{2-6}
 & IO & $+$ & \multicolumn{3}{|c|}{excluded}
\\ \cline{3-6}
 &  & $-$ & \multicolumn{3}{|c|}{excluded}
\\ \hline\hline
  & NO & $+$ & $\overline{1.1^\circ} \sim 12.5^\circ$ & $\underline{38.6^\circ} \sim \overline{53.1^\circ}$ & $0.0\pm 0.047$
\\ \cline{3-6}
New &  & $-$ & \multicolumn{3}{|c|}{excluded}
\\ \cline{2-6}
 & IO & $+$ & \multicolumn{3}{|c|}{excluded}
\\ \cline{3-6}
  & & $+$ & $\overline{1.1^\circ} \sim 12.5^\circ$ & $\underline{38.6^\circ} \sim \overline{53.1^\circ}$ & $0.0\pm 0.047$
\\ \hline
\end{tabular}
\end{center}
\caption{The allowed ranges of $\theta_{13}$, $\theta_{23}$, and $J_{\rm CP}$ with respect to the $3\sigma$ constraints of $\Delta m_{31}^2$, $\Delta m_{21}^2$, and $\theta_{12}$ given in Eq. (\ref{eq:gfit}),
where the ranges of $\theta_{23}$ are also restricted to be within the $3\sigma$ bounds.
The under and over lines to the values represent the lower and upper edges of $\theta_{23}$, respectively, while
the values surrounded by $\langle\rangle$ are limited by the $3\sigma$ upper bound of $\theta_{12}^{}$.
} \label{tab:smry}
\label{tab:}
\end{table}
In Table \ref{tab:smry}, we summarize the allowed ranges of $\theta_{13}^{}$, $\theta_{23}^{}$, and $J_{\rm CP}^{}$ with respect to not only the $3\sigma$ constraints of $\Delta m_{31}^2$, $\Delta m_{21}^2$, and $\theta_{12}$  from Eq. (\ref{eq:gfit}) but also that of $\theta_{23}^{}$. 
Note that the maximum values of $\theta_{13}$ and $J_{\rm CP}^{}$ in the new mixing case correspond to 
both
the $3\sigma$ edges of $\theta_{23}^{}$ and
the assumed maximal value of $I^{\rm loop}_{}$, so that
 they can be larger with an even more larger value of $I^{\rm loop}_{}$.

\section{simple realization of finite quantum corrections}
\begin{table}[h]
\begin{center}
\begin{tabular}{|c||c|c|c|c|c|c|c|c|}\hline
 & $Q_L^{}$ & $d_R^{}$ & $u_R^{}$ & $L_L^{}$ & $\ell_R^{}$ & $H_u^{}$ & $H_d^{}$ & $\Delta$ \\ \hline
$SU(2)_L^{}$ & $2$ & $1$ & $1$ & $2$ & $1$ & $2$ & $2$ & $3$ \\ \hline
$U(1)_Y^{}$ & $1/3$ & $-2/3$ & $4/3$ & $-1$ & $-2$ & $1$ & $1$    & $2$ \\ \hline
$Z_4^{}$ & $0$ & $1$ & $1$ & $0$ & $1$ & $1$ & $3$ & $0$ \\ \hline
\end{tabular}
\end{center}
\caption{The particle content with charge assignments of the model.}
\label{tab:model}
\end{table}
We show a simple realization of the finite quantum correction assumed in Eq. (\ref{eq:dM}). 
We consider a two-Higgs-doublet-extension of the SM and further introduce an $SU(2)_L^{}$ triplet scalar, $\Delta$, which possesses $Y=2$.
Besides, we impose a $Z_4^{}$ symmetry in order to avoid the dangerous flavor changing neutral currents in the quark sector.
The particle content with charge assignments of the model is summarized in Table \ref{tab:model}. 
The Yukawa Lagrangian and scalar potential are given by
\begin{eqnarray}
{\cal L}_y^{} = 
  Y_d^{} ~\overline{Q}_L^{} H_d^{} d_R^{} 
+ Y_u^{} ~\overline{Q}_L^{} (i\sigma_2^{} H^{*}_u) u_R^{}
+ Y_\ell^{} ~\overline{L}_L^{} H_d^{} \ell_R^{} 
+ Y_\Delta^{} ~L_L^T C (i\sigma_2^{} \Delta) L_L^{}+ h.c.\ ,
\end{eqnarray}
where $\sigma_2^{}$ is the Pauli matrix and $C$ stands for the charge conjugation matrix, and
\begin{eqnarray}
V
&=&
  n_u^2 H_u^\dag H_u^{} + n_d^2 H_d^\dag H_d^{}
+ n_\Delta^2 \Tr[\Delta\Delta^\dag]
+ \mu\left[ H_u^T (i\sigma_2^{} \Delta^\dag) H_d^{} + h.c. \right]
\hspace{2.6cm}\nonumber \\
&&
+ \lambda_1^{} |H_u^\dag H_u^{}|^2 
+ \lambda_2^{} |H_d^\dag H_d^{}|^2
+ \lambda_3^{} \left[ (H_u^\dag H_d^{})^2 + h.c. \right]
+ \lambda_4^{} (H_u^\dag H_d^{})(H_d^\dag H_u^{})\nonumber \\
&&
+ \lambda_5^{} (H_u^\dag H_u^{})(H_d^\dag H_d^{})
+ \lambda_6^{} \left(\Tr[\Delta\Delta^\dag]\right)^2
+ \lambda_7^{} \Tr(\Delta\Delta^\dag \Delta\Delta^\dag) 
\nonumber\\
&&
+ \lambda_8^{} (H_u^\dag H_u^{})\Tr[\Delta\Delta^\dag] 
+ \lambda_9^{} (H_d^\dag H_d^{})\Tr[\Delta\Delta^\dag]\nonumber \\
&&
+ \lambda_{10}^{} ~H_u^\dag \Delta\Delta^\dag H_u^{}
+ \lambda_{11}^{} ~H_d^\dag \Delta\Delta^\dag H_d^{} ,
\end{eqnarray}
respectively, with the following conventions of the scalars:
\begin{eqnarray}
H_{u,d}^{}=
\bmx{c}
 \phi^+_{u,d} \\ h_{u,d}^{} + i \eta_{u,d}^{}
\emx,\ \ 
\Delta=
\bmx{cc}
\frac{1}{\sqrt{2}}\Delta^+ & \Delta^{++} \\
\Delta^{0}+i\delta & -\frac{1}{\sqrt{2}}\Delta^+
\emx .
\end{eqnarray}
In the potential, all the couplings are chosen to be real without  loss of generality.

Although there are many parameters in the potential, not all of them are indispensable for the following discussions.
Hence, just for simplicity, we turn off $\lambda_4^{} \cdots \lambda_{11}^{}$ from now on.
By solving the stationary conditions for $h_{u,d}^{}$ and $\Delta^0$, we arrive at the VEV configurations:
\begin{eqnarray}
v_u^2=
\frac{\mu v_\Delta^{} \tan^{-1}\beta-n_u^2}
{2\lambda_1+2\lambda_3\tan^{-2}\beta},~~
v_d^2=
\frac{\mu v_\Delta^{} \tan\beta-n_d^2}
{2\lambda_2+2\lambda_3\tan^2\beta},~~
v_\Delta^{}=
\frac{\mu v_u v_d}{n^2_\Delta}~, \label{eq:vevs}
\end{eqnarray}
where $\tan\beta=v_u/v_d$. 

As we shall explain later, $v_\Delta$ is responsible for the tree-level neutrino masses and thus, the smallness of neutrino masses originates in that of $\mu/n_\Delta^2$.
In general, both $\mu$ and $n_\Delta^{}$ can take extremely large values.
Nevertheless, in order to make the discussion more simple, we restrict ourselves to the case of $n_\Delta^{} \gg v_{u,d}^{} \gg \mu \gg v_\Delta^{}$. 
In this limit, the mass eigenstates of singly-charged scalars are given by
\begin{eqnarray}
P^\pm_{} = \cos\beta ~\phi^\pm_u - \sin\beta ~\phi^\pm_d,~~
G^\pm_{} = \sin\beta ~\phi^\pm_u + \cos\beta ~\phi^\pm_d
\end{eqnarray}
with their masses $M^2_{P^\pm_{}} = 2\lambda_3 v^2_{}$ and $M^2_{G^\pm_{}} = 0$, where $v^2_{}=v^2_u + v^2_d = (174~\gev)^2$.
Note that the mass scales of $\Delta^{\pm\pm}$,  $\Delta^\pm$, $\Delta^0$, and $\delta$ are mutually described by $n^{}_\Delta$.

\begin{figure}[t]
\begin{center}
\includegraphics*[width=0.6\textwidth]{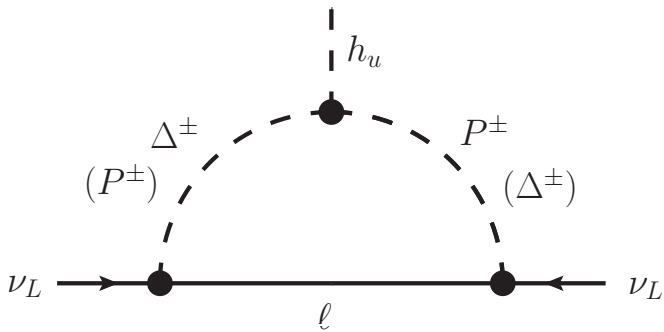}
\caption{\footnotesize A one-loop neutrino mass operator. }\label{fig:1loop}
\end{center}
\end{figure}
The charged fermions acquire their masses through the Higgs mechanism, given by 
\begin{eqnarray}
M_d^{}=Y_d^{} v_d^{}, ~~M_u^{}=Y_u^{} v_u^{}, ~~M_\ell^{}=Y_\ell^{} v_d^{}, 
\end{eqnarray}
while Majorana neutrino masses result from the $Y_\Delta L^T \Delta L$ term after $\Delta^0$ develops a VEV \cite{type2},  given by
\begin{eqnarray}
M_\nu^0 = Y_\Delta v_\Delta 
= Y_\Delta\frac{\mu v_u v_d}{n_\Delta^2}.
\label{eq:0mn}
\end{eqnarray}
Furthermore, from $Y_\Delta L^T \Delta L$, $Y_\ell \overline{L} H_d \ell$, and $\mu H_u^T \Delta^\dag_{} H_d^{}$ terms, 
a one-loop neutrino mass operator can be induced via the diagram depicted in Fig. \ref{fig:1loop}.
If we require $\tan\beta \gg 1$, the one-loop mass matrix can be written as 
\begin{eqnarray}
\delta M_\nu^{} \simeq 
\frac{M_\nu^0 D_\ell^2 + D_\ell^2 M_\nu^0}{v_{}^2}
\times \left(-\frac{\tan^2\beta}{16\pi^2}
\frac{1}{1 - M_{P^\pm}^2/M_{\Delta^\pm}^2}
\ln\frac{M_{P^\pm}^2}{M_{\Delta^\pm}^2}\right), 
\end{eqnarray}
where we have assumed $\sin\beta = 1$. 
For example, by taking $M_{P^\pm}=10^2~\gev$, $M_{\Delta^\pm}=10^3~\gev$ (or $10^5~\gev$), and $\tan\beta=32$ (or $38$), we obtain $I^{\rm loop}\simeq 30$ (or $126$) with $\mu\simeq 10^{-6}~\gev$ (or $10^{-2}~\gev$).

\section{Summary}
We have applied the scheme of finite quantum corrections to the TBM, BM, and DC mixing patterns and systematically investigated how large $\theta_{13}^{}$ can depart from $0^\circ$.
We have found that (i) $\theta_{13}$ can largely deviate from $0^\circ$ with a relatively large $I^{\rm loop}$, (ii) the large deviation of $\theta_{13}$ is always accompanied with a large deviation of $\theta_{23}$ from its initial value, and (iii) in the large $I^{\rm loop}$ region, $\theta_{12}$ cannot always  drastically depart from its initial value.
Because of these features, unfortunately, all the TBM, BM, and DC patterns cannot reproduce the experimentally favored mixing angles and masses after taking the finite quantum corrections into account.
Instead of these well known mixing patters, we have shown an example of a new tree-level mixing matrix, which works very well with our scheme.
We have also proposed a simple realization of the finite quantum corrections by introducing new $SU(2)_L^{}$ doublet and triplet scalars with a $Z_4^{}$ symmetry.
Finally, we remark that the above conclusions may be valid  only for the finite quantum corrections introduced in Eq. (\ref{eq:dM}).
Different types\footnote{
In Ref. \cite{AGX}, two-loop finite quantum corrections are also discussed.
} 
of the corrections may result in different conclusions.
Nevertheless, we do not go into more detail on this possibility since it goes beyond the scope of this paper.

\begin{acknowledgments}
We are grateful to Z.Z. Xing for useful discussions and the early stage of this work.
The work of T.A. was supported in part by the National Natural Science Foundation of China under Grant No. 10875131. 
The work of C.Q.G. was partially supported by the National Science Council under Grant No. NSC-98-2112-M-007-008-MY3 and  National Center of
Theoretical Science.
\end{acknowledgments}

\end{document}